\documentclass[11pt]{article}
 \pdfoutput=1
\usepackage{amsmath}
\usepackage{amssymb}
\usepackage{graphicx}
\usepackage{color}
\usepackage{authblk}
\usepackage[margin=1in]{geometry}

\title{\textbf{Light-induced swirling and locomotion }}
\author[1]{Ameneh Maghsoodi}
\author[1,*]{Kaushik Bhattacharya}

\affil{Division of Engineering and Applied Science, California Institute of Technology, Pasadena, CA 91125}
\affil[*]{Corresponding author: bhatta@caltech.edu}

\date{\today}

\begin{document}
\maketitle

\paragraph{Abstract}
Photomechanical liquid crystal elastomers (LCEs) are responsive polymers that can convert light directly into mechanical deformation. This unique feature  makes these materials an attractive candidate for soft actuators capable of remote and multi-mode actuation. In this work, we propose a three-dimensional multi-scale model of the nonlinear and nonlocal dynamics of fibers of photomechanical LCEs under illumination.  We use the model to show that a pre-stressed helix-like fibers immersed in a fluid can undergo a periodic whirling motion under steady illumination.  We analyze the photo-driven spatiotemporal pattern and stability of the whirling deformation, and provide a parametric study.     Unlike previous work on photo-driven periodic motion, this whirling motion does not exploit instabilities in the form of snap-through phenomena, or gravity as in rolling.   We then show that such motion can be exploited in developing remote controlled bio-inspired microswimmers  and novel micromixers.


\paragraph{Significance}
The design of untethered devices with no or limited onboard power that can be activated remotely remains a continuing challenge in soft robotics.   This work addresses this challenge by developing a method of generating a whirling motion in a pre-stressed photomechanical liquid crystal elastomer fiber using  steady illumination, and then showing how this motion may be exploited for applications such as microswimmers and micromixers.  More broadly, this provides an unusual example of a physical system capable of periodic motion under steady stimulus that does not exploit instabilities.
\section{Introduction}
Liquid crystal elastomers (LCEs) are rubbery networks composed crosslinked polymer chains that contain liquid-crystalline mesogens in their main or in side chains. External stimuli like heat changes the ordering of the liquid-crystalline (LC) mesogens leading to a change in shape or stimuli-induced deformation.  This phenomenon is reversible.   LCEs containing light-sensitive molecules, such as azobenzene (azo) photochromes, exhibit a reversible photomechanical behavior: they absorb light energy and convert it into mechanical energy by changing their shape. This fascinating photomechanical effect arises from the {\it trans-cis} isomerization of azobenzene dyes, a process in which the azo molecules absorb light energy and change their conformation from a linear {\it trans} isomer to a bent {\it cis} isomer.   The steric interaction between the azo and LC molecules consequently changes the LC ordering leading to a photo-induced shape-change.  Upon removing the light, the {\it cis} isomers thermally relax to the {\it trans} state leading to a reversal of the photo-induced shape-change.

Recent works demonstrate the potential of photomechanical materials for soft robotics because of their remarkable features: 
1) properly synthesized photomechanical materials can undergo large, reversible deformation upon light irradiation,  
2) these materials are lightweight and soft, and hence, appropriate to generate flexible motion in soft robotics, 
3) light is a clean power source that can actuate a photomechanical object from distance, thereby eliminating the need for on-board power or tethers, and 
4) light-induced deformations can be controlled by changing light intensity, wavelength, or polarization \cite{kumar2016chaotic,zeng2018light,tabiryan2005polymer} thereby providing a platform for multiplexing and control.  

\begin{figure}
\begin{center}
  \includegraphics[width=4in]{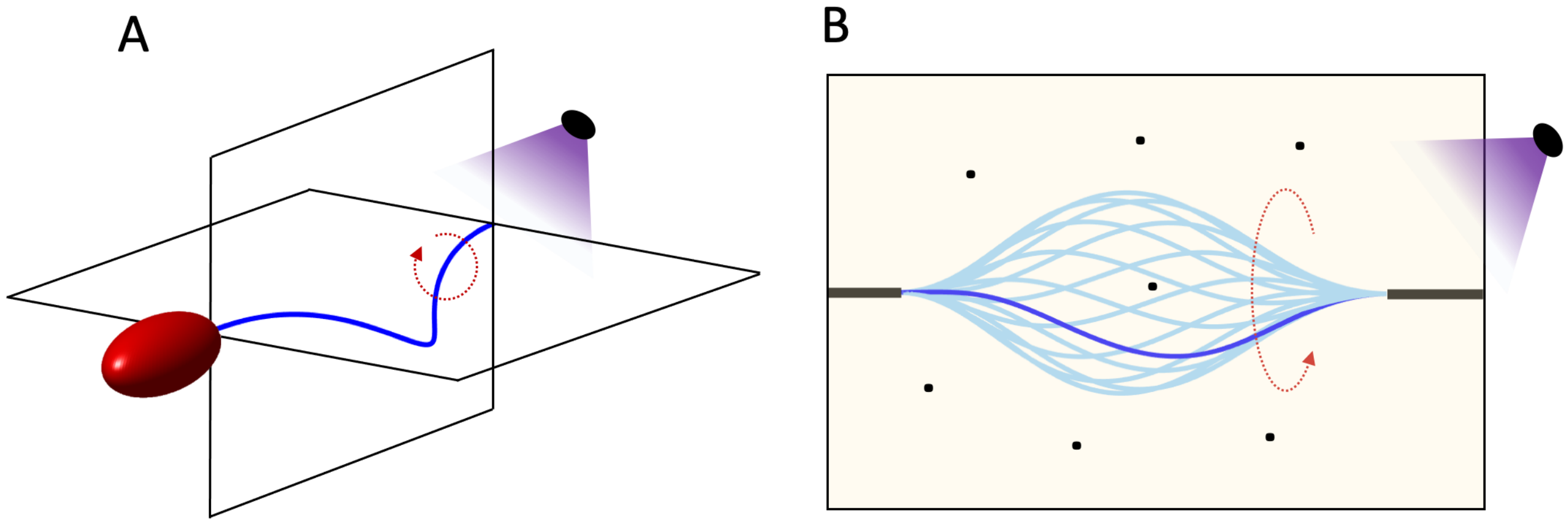}
  \caption{Potential applications of photo-driven periodic motion of fibers: microswimmers and micromixers.}
  \label{fig0}
\end{center}
\end{figure}

In particular, recent attention has focussed on locomotion in soft-robotics.  A long-standing problem in the use of active materials is the need to reset the material after each actuation stroke.  The ability to rapidly turn on and turn off  light enables such a reset.  This has been exploited to build legged microwalkers \cite{zeng2015light},  leg-free caterpillar-like crawling \cite{Zeng2018LightDrivenCM} and swimming flagellum \cite{huang2015miniaturized}.  Even more promising, the coupling between photo-isomerization, light propagation and nonlinear mechanics can be exploited to generate cyclic motion even under steady illumination.  Yamada {\it et al.} \cite{yamada2008photomobile} demonstrated a ring can be made to roll by simultaneously illuminating it with a combination of UV and natural light at two different but carefully chosen locations.    Wie {\it et al.} \cite{wie2016photomotility} demonstrated that a flat monolithic polymer films made of azo-containing liquid crystalline polymer networks changes its conformation to a spiral ribbon under illumination. The ribbon also translates a large distance with continuous irradiation.   Gelebart {\it et al.} \cite{gelebart2017making} created a wave like motion a doubly clamped azo-LCE strip using steady illumination and exploited this for a framed walker under steady illumination.  In  these examples, the motion is generated by gravity and the reaction from the surface.   In this work, we explore cyclic motion under steady illumination in an immersive environment, and propose potential applications as bio-inspired micro-swimmers and micro-mixers as shown in Figure \ref{fig0}.

The generation of cyclic deformation has also been a subject of interest in chemo-mechanical systems since the demonstration of oscillatory swelling and deswelling in hydrogels \cite{yoshida,crooks}.  In these systems, cyclic chemistry like the Belousov-Zhabotinsky reactions or Landolt pH oscillators are combined with hydrogels to obtain a chemo-mechanical oscillations.   These have recently been used to demonstrate spatio-temporal oscillatory motion that can be exploited for peristaltic motion and walking \cite{kim}, as well as homeostatic feedback loops \cite{he}.  A theoretical model of these systems show that the system is statically bistable and the dissipative kinetics leads to a these chemo-mechanical oscillations.  In this way, these systems are conceptually similar to the  wave like motion \cite{gelebart2017making}.  In this work, we explore the emergence of periodic behavior in a stable system.

We begin by developing a three-dimensional theory of photomechanical rods.   The general theory of photomechanical coupling was developed by Corbett and Warner \cite{corbett_nonlinear_2006}, and used by Warner and Mahadevan \cite{warner2004photoinduced} to study the development of light-induced spontaneous curvature in beams and plates under the assumption of shallow penetration.  Corbett {\it et al.} \cite{corbett2015deep} studied deep penetration, and the deformation of beams under stress-free conditions (i.e., without loads).   Korner {\it et al.} \cite{korner_nonlinear_2020} developed the theory of photomechanical beams under loads and used it to explain the rolling of rings \cite{yamada2008photomobile} and wave-like motion of doubly-clamped beams \cite{gelebart2017making}.  
All of this work was limited to the two-dimensional setting of beams.  In this paper, we extend these ideas to the three dimensional setting of rods by combining the photomechanical actuation with  nonlinear Kirchhoff rod theory.  We do so in a dynamical setting with inertia. 

We then use the theory to exploit photomechanical coupling and pre-stressed structural configuration to generate periodic {\it whirling} motion under steady illumination; a fascinating photomechanical response that, to our knowledge, has not been reported in literature. Importantly, we show that such whirling motion of a fiber immersed in a viscous fluid generates propulsion along its rotational axis.   This can be exploited in developing remote controlled bio-inspired microswimmers and micromixers (Figure \ref{fig0}).  We also show that this phenomenon does not rely on instabilities of flapping, or on gravity.  Finally, we show that we can control the photoresponse using illumination conditions (light intensity and direction), pre-stressed configuration of fiber (twist and bending) and cross-sectional shape (circular and rectangular).  These provide opportunities to design microactuators for different conditions, and for control. Together, the model and the results  open a new avenue for the development of remote actuated and controlled soft actuators. 

%
\section{Model}
\begin{figure}[t]
\begin{center}
  \includegraphics[width=4in]{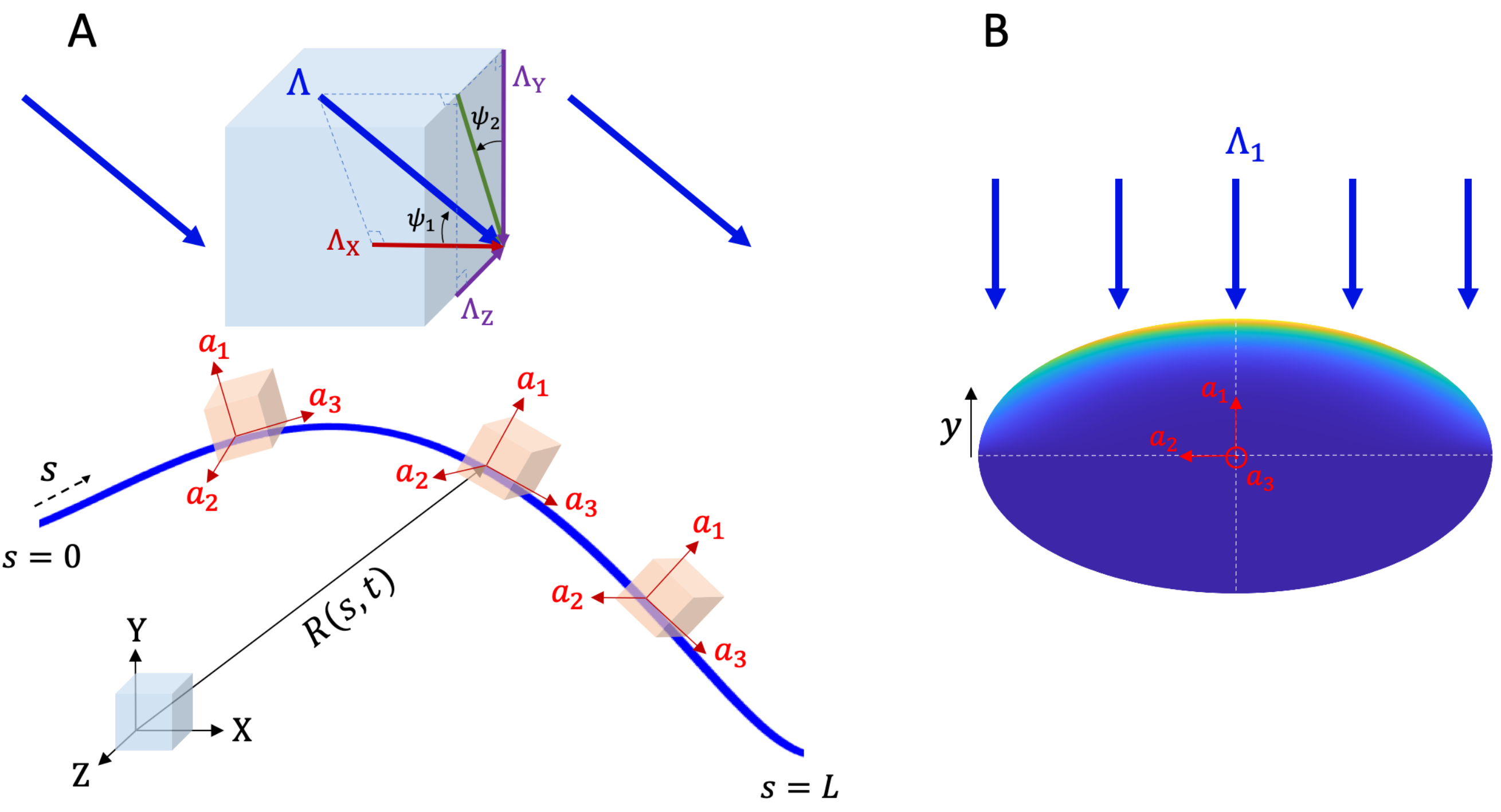}
  \caption{Dynamic model of a photomechanical fiber under steady illumination. (A) Nonlinear dynamics of the fiber is described using Kirchhof rod theory. Frame XYZ represents the fixed inertial frame and $\{\hat a_i \}$ represents the body-fixed frame at each cross-section of the fiber. The steady illumination is defined by the intensity $\Lambda$ and angles $\psi_1$ and $\psi_2$ with respect to X axis and plane XY, respectively. (B) Schematic of the cross-section of fiber under illumination along axis $a_1$.}
  \label{fig1}
\end{center}
\end{figure}

Consider a long, thin azo-LCE fiber subjected to a steady illumination $\vec \Lambda$ that makes angles $\psi_1$ and $\psi_2$ with fixed frame XYZ as illustrated in Figure \ref{fig1}A.   We describe the centerline of fiber as a time-dependent space curve $\vec R(s,t)$ where $s$ is the arc-length along the fiber and $t$ is time.   Let $\{\hat a_i(s,t)\}$ be a body-fixed orthonomal frame at point $s$ and time $t$ with unit vectors $\hat a_1$ and $\hat a_2$ coinciding with the principal axes of the cross-section and $\hat a_3$ normal to the cross-section.  We assume that the fiber is inextensible and unshearable so that $\hat a_3$ is always tangent to the space curve.  We define the curvature vector $\vec \kappa = \{ \hat a_3' \cdot a_1,  \hat a_3' \cdot a_2, \hat a_1' \cdot \hat a_2\}$ in the $\{\hat a_i \}$ frame.  Note that the rod may have intrinsic (stress-free, illumination free) curvature.

We now specify the photomechanical constitutive behavior of the rod building on the ideas of   \cite{corbett_nonlinear_2006, korner_nonlinear_2020}.  Consider a short section of the filament at position $s$ and time $t$.  Suppose for the moment that it is subjected to an illumination along axis $a_1$ with intensity $\Lambda_1$ as shown in Figure \ref{fig1}B.  As the light diffracts into the cross-section, it is absorbed by the photochrome molecules and attenuates.  In this work, we assume shallow penetration governed by Beer's law.  Recall that, according to Beer's law, when a light impinges on a flat surface with intensity $\Lambda_0$, the intensity $\Lambda(d)$ at a depth $d$ is given by $\Lambda_0 \exp(-d/d_0)$ where $d_0$ is the penetration depth.  Generalizing this to our general cross-section, the intensity at any point $\vec y$ on the cross-section is given by $\Lambda(\vec y) = \Lambda_1 f_1(\vec y)$ for some function $f_1: A \to [0,1]$ which quickly decays to zero away from the surface as illustrated in Figure \ref{fig1}B.  Examples of this function are provided in the supplementary material.

The photochromes in the {\it trans} state absorb light and transform into the {\it cis} state which can thermally relax back to the {\it trans} state.  Therefore, the number density $n_c (\vec y)$ of {\it cis} molecules at position $\vec y$ evolves according to 
\begin{equation}
\tau \dot n_c(\vec y,t) = -n_c (\vec y,t) +\alpha \Lambda(\vec y)  = -n_c +\alpha \Lambda_1 f_1(\vec y) \label{eq:a0},
\end{equation}
where the dot represents the material time derivative, $\tau$ is thermal relaxation time (or $cis$ lifetime), and $\alpha$ is a constant depending on material properties and forward-backward reaction rates.  This in turn introduces a spontaneous axial strain $\varepsilon_c = - \lambda n_c$ where $\lambda$ is a constant of proportionality with $\lambda>0$ corresponds to a contraction. It follows that the axial normal stress 
$\sigma (\vec y) = E [(\kappa_2 - \kappa_2^{in})\vec y \cdot a_1 + \lambda n_c (\vec y)]$ where $\kappa_2 = \vec \kappa \cdot a_2$, $\kappa_2^{in}$ is the intrinsic (stress-free, illumination-free) curvature in the 2-direction, and $E$ is the Young's modulus. The bending moment along the $\hat a_2$ is 
\begin{equation} \label{eq:q1}
q_2 = \int_A (\vec y \cdot a_1) \sigma (\vec y) \mathrm{d}A = B_2 (\kappa_2 - \kappa_2^0), \quad \kappa_2^0=\kappa_2^{in} - \int_A \frac{\lambda n_c (\vec y)}{I_2}(\vec y \cdot a_1) \mathrm{d}A
\end{equation}
where $B_2 = E I_2$ is the bending rigidity along $\hat a_2$ and $I_2$ is the second moment of inertia along $\hat a_2$.  Now, differentiating $\kappa_2^0$ with respect to time and using (\ref{eq:a0}), we conclude that 
\begin{equation}
\tau \dot \kappa_2^0  + (\kappa^0_2 - \kappa^{in}_2) = \gamma_1 \Lambda_1, \quad \gamma_1 = \frac{-\alpha \lambda}{I_2} \int_A f_1 (\vec y) (\vec y \cdot a_1) \mathrm{d}A.
\end{equation}

In the case of general illumination, we split it into components along $\hat a_1$ and $\hat a_2$ to conclude that the photomechanical constitutive behavior of the beam is given by
\begin{eqnarray}
\vec q = {\mathbf B} (\vec \kappa - \vec \kappa^0),  \label{eq:const} \\ 
\tau  \dot{\vec \kappa}^0  - (\vec \kappa^0 - \vec \kappa^{in}) = {\mathbf \Gamma} \vec \Lambda  \label{eq:rate}
\end{eqnarray}
where $\vec q$ is the vector of bending moments and torque, ${\mathbf B}$ is the tensorial rigidity (diag$[EI_1,EI_2,GJ]$ in the $\{\hat a_i\}$ basis with bending and torsional rigidities and ${\mathbf \Gamma} = \gamma_1 \hat a_1 \otimes \hat a_1 + \gamma_2 \hat a_2 \otimes \hat a_2$ is the photomechanical rate tensor (diag$[\gamma_1, \gamma_2,0]$ in the $\{\hat a_i\}$ basis).

Finally, we complete the system of equations by considering the balance of linear and angular momenta \cite{dill1992kirchhoff,goyal2005nonlinear},
\begin{eqnarray} 
\left(\frac{\partial \vec f}{\partial s}+\vec\kappa \times \vec f\right) +\vec f_{ext}=m\left(\frac{\partial \vec v}{\partial t}+\vec \omega \times \vec v\right)  \label{eq:1},\\
\left(\frac{\partial \vec q}{\partial s}+\vec \kappa \times \vec q\right)+\hat {a}_3 \times \vec f +\vec q_{ext}=\left(\mathbf I_m\frac{\partial \vec \omega}{\partial t}+\vec \omega \times \mathbf I_m \vec \omega\right)\label{eq:2}
\end{eqnarray} 
where $\vec v$ is the particle velocity, $\partial \over \partial t$ is the spatial time derivative, $\vec \omega$ the angular velocity of the frame $\{\hat a_i\}$, and $\vec f_{ext}$ and $\vec q_{ext}$ are the external or applied force and moment per unit length respectively;
the inextensibility 
\begin{eqnarray} 
\frac{\partial \vec v}{\partial s}+\vec \kappa \times \vec v=\vec \omega \times \hat {a}_3 \label{eq:3};
\end{eqnarray} 
 the compatibility between curvature and angular velocity
\begin{eqnarray} 
\frac{\partial \vec \omega}{\partial s}+\vec \kappa \times \vec \omega=\frac{\partial \vec \kappa}{\partial t} \label{eq:4};
\end{eqnarray} 
and appropriate initial and boundary conditions. 

If the fiber is immersed in a fluid, it experiences a hydrodynamic drag force per unit length that acts as the external force in (\ref{eq:1}):
\begin{equation} 
\vec f_{ext} = \vec f_{drag} =  -\mathrm {diag}(C_1,C_2,C_3) \vec v \label{eq:10};
\end{equation}
where the normal, bi-normal, and tangential drag coefficients are estimated to be $C_1=C_2=4\pi\mu/(\mathrm{ln}(L/D)+0.84)$ and $C_3=2\pi\mu/(\mathrm{ln}(L/D)-0.2)$ for a circular cross-section, respectively, in the limit of low Reynolds number \cite{howard2001mechanics}. Above, $\mu$ is the dynamic viscosity of fluid, and $L$ and $D$ are the length and diameter of fiber, respectively. The drag force is incorporated into the dynamic model through the term $\vec f_{ext}$ in (\ref{eq:1}).

To solve the nonlinear initial-boundary-value problem above, we employ a finite difference method in both space and time to discretize the equations. Starting with the initial condition of fiber at $t=0$, the discretized equations are integrated over space at each successive time step. During spatial integration, the boundary conditions are satisfied using the shooting method for boundary-value problems. For details on the numerical method, see \cite{gatti2002physical,maghsoodi2016first,gobat2002generalized}.

\section{Photo-driven motion of a buckled and twisted fiber}
We now use the model to study the deformation of initially straight ($\kappa^{in} = 0$)  fibers with circular ($\gamma_1 = \gamma_2, I_1 = I_2$ ) cross-section that are buckled and possibly twisted, clamped at the two ends and illuminated. We use the parameters listed in Table \ref{table:1} unless otherwise specified.     A detailed parameter study is provided in supplementary information.

\begin{table}[ht]
\caption{Simulation parameters} 
\centering 
\begin{tabular}{l l } 
\hline\hline 
Parameter (symbol) &  Value \\ [0.5ex] 
\hline 
Young's modulus ($E$)   &   4 GPa \cite{smith2014designing}  \\ 
Fiber diameter ($D$)    &   15 $mu$ m \cite{smith2014designing} \\
Fiber length ($L$)   &   15 mm \cite{smith2014designing} \\
Penetration depth ($d$)   &   0.56 $\mu$ m \cite{korner_nonlinear_2020} \\
Relaxation time ($\tau$)   &   0.1 s \\
{\it cis}-strain proportionality ($\lambda$)  &  0.05 \\ [1ex] 
\hline 
\end{tabular}
\label{table:1} 
\end{table}

\begin{figure}
\begin{center}
  \includegraphics[scale=0.7]{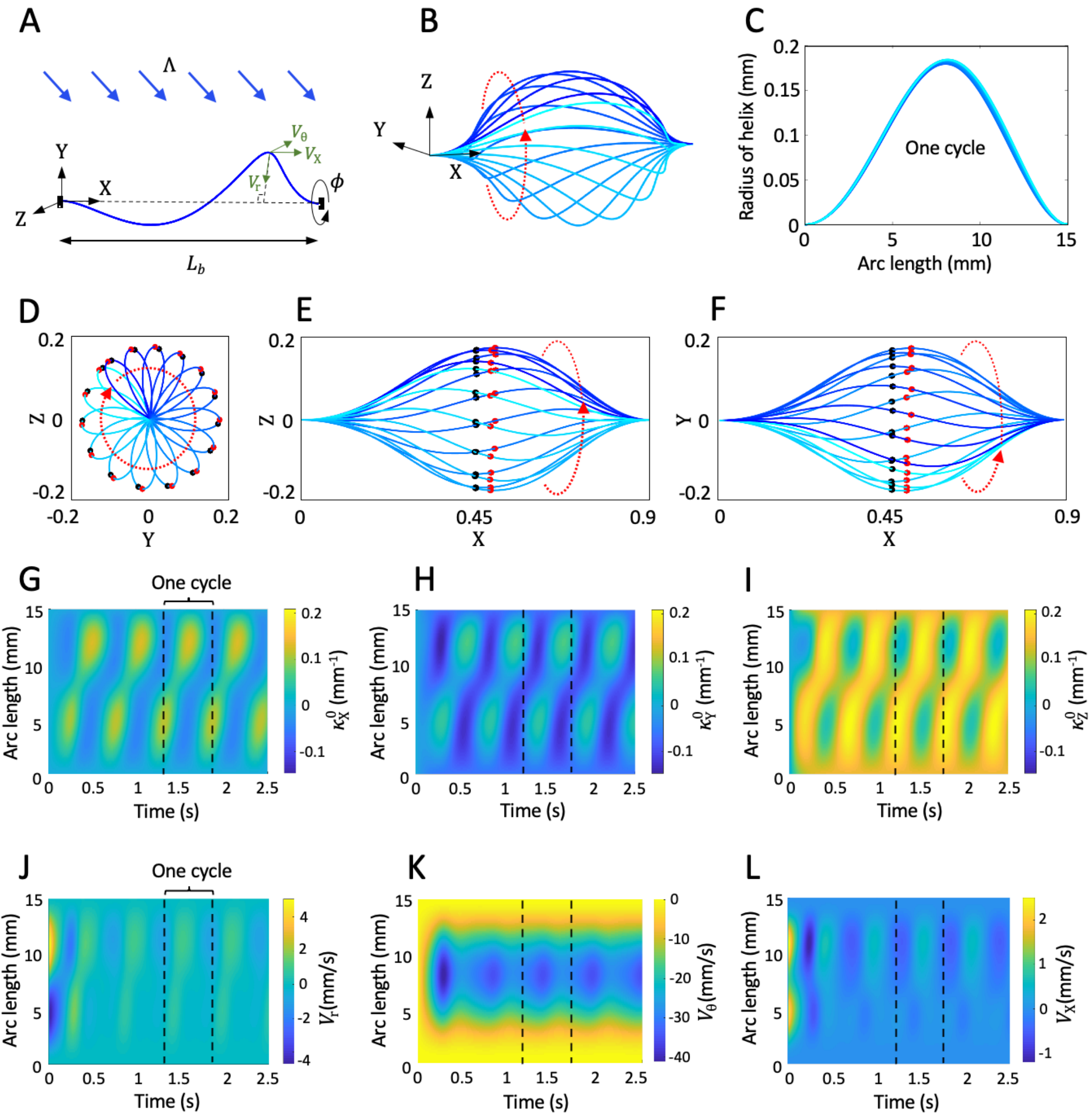}
  \caption{Light-induced periodic motion of a twisted-buckled fiber under steady illumination. (A) Schematic of fiber under illumination. (B) The buckled-twisted fiber undergoes a periodic whirling; dark blue refers to start of the cycle and light blue indicates the end of the cycle. (C) Radial distance of fiber relative to the X axis during one cycle. (D-F) Three orthogonal views of the photoactivated fiber. (G-I) Evolution of the light-induced spontaneous curvature $\kappa_0$ of the fiber in XYZ frame. Red dots refer to maximum radius of fiber and black dots refer to mid-point. (J-L) Evolution of velocity of the fiber in cylindrical coordinate $r\theta X$ shown in (A). In this simulation, $L_b=0.9L$, $\phi=60^{\circ}$, $\Lambda=300 \mathrm{W/m^2}$, $\psi_1=30^{\circ}$, and $\psi_2=20^{\circ}$.}
  \label{fig2}
\end{center}
\end{figure}
We begin by defining a pre-stressed doubly clamped conformation as the initial state of the fiber. Assume that the fiber is initially straight ($\vec\kappa_{in}=0$) with length $L$ and the nematic director is along the axis of the fiber (so $\lambda >0$ in Eq. (\ref{eq:q1})). We compress and twist the fiber quasi-statically by applying a compressive force along and a torque about the axis of fiber till the fiber buckles (the final end to end distance is $L_b < L$ and angle of twist is $\phi$).  We then clamp the two ends by constraining their position and rotation till the fiber is in equilibrium. This is initial state of fiber which we then illuminate as shown in Figure \ref{fig2}A.

\paragraph{Typical result}
A typical result ($L_b = 0.9L, \phi = 60^\circ, \Lambda =  300 \text{W/m}^2, \psi_1 = 30^\circ, \psi_2 = 20^\circ$) is shown in Figure \ref{fig2}B-L.  After an initial transient, the fiber undergoes a steady cyclic motion, snapshots of which over one cycle are displayed in Figure \ref{fig2}B (in perspective view) and D-F (three normal projections).  Note from Figure \ref{fig2}D that the rotation about the X axis (line joining the end points) is uniform and from Figure \ref{fig2}E,F that the shape is slightly asymmetric  about the X-axis and about the center.  Snapshots of the radial distance (distance to the line joining the end points)  during the cycle is shown in Figure \ref{fig2}C; note that this is almost invariant.  In short, the buckled and twisted fiber rotates about the axis quite uniformly and with an almost a rigid shape.  However, the rotation is not rigid since the ends are clamped.
Figure \ref{fig2}G-I plot the spatio-temporal evolution of the spontaneous curvature of the fiber (components of $\vec\kappa_0$ with respect to the laboratory frame) over first five cycles, while Figure \ref{fig2}J-L plot the velocity. These confirm that the motion is indeed periodic after an initial transient.  We also note that after an initial transient, the radial and axial components of velocity remain very small confirming the almost fixed shape we observed earlier.  The angular velocity shows small periodic fluctuations showing that the rotational motion is almost, but perfectly, uniform.

As the fiber is illuminated, the photomechanical coupling causes a change in spontaneous curvature which leads to a change of shape.  The overall length of the fiber endows it with a very small twisting stiffness, and this enables the fiber to accommodate this change of shape by a rotary motion.  This changes the illumination conditions (the angle between the direction of light propagation and the tangent to the fiber) leading to a further change of shape and this cascades into a cyclic motion that we observe.   We emphasize that  the fiber is clamped at both ends, so it is not free to rotate about the X-axis.  This is consistent with the periodic fluctuations in the angular velocity. Still, the small twisting stiffness makes this a low energy manifold of deformation, thereby enabling the motion.  Importantly, there is no instability which we discuss further in the case with zero twist.

\begin{figure}[t!]
\begin{center}
  \includegraphics[width=6.5in]{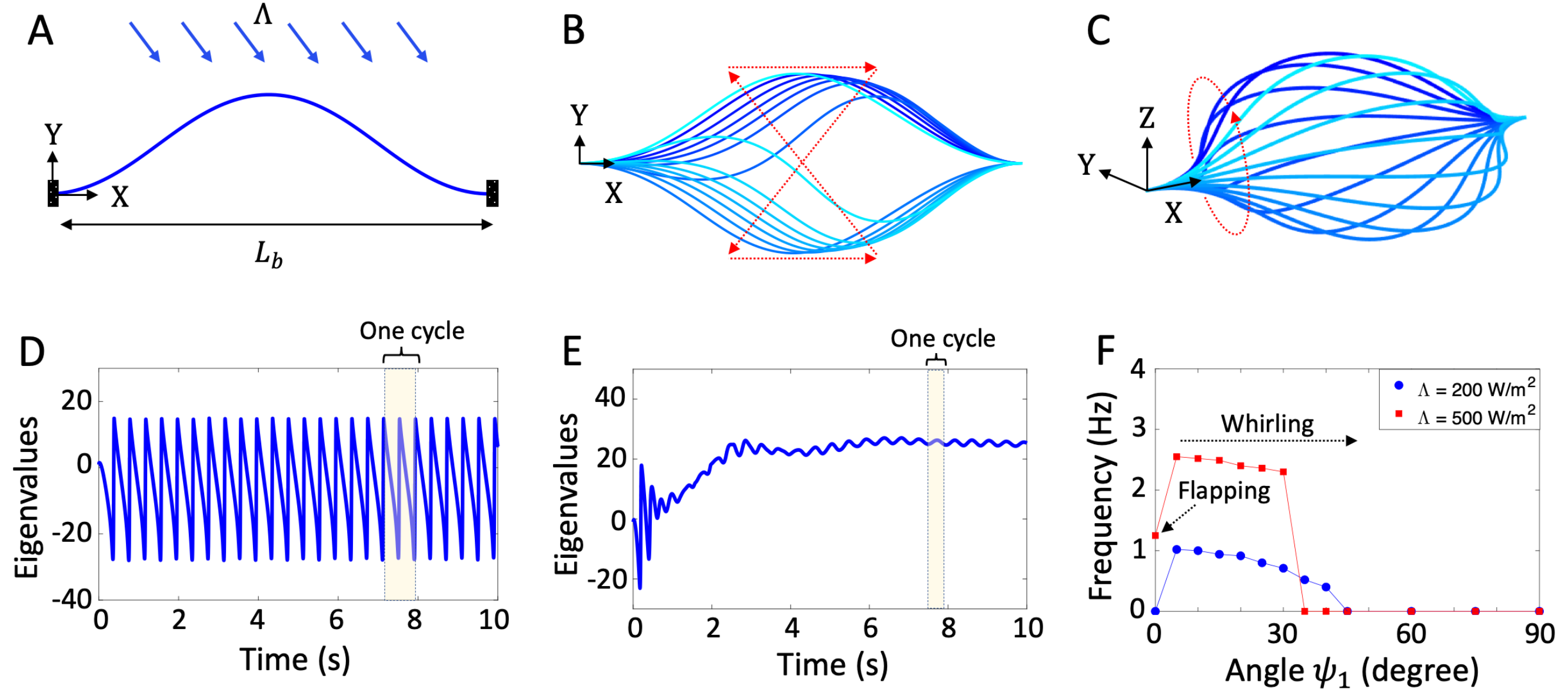}
  \caption{Light-induced periodic motion of a buckled fiber under steady illumination. (A) Pre-stressed buckled fiber is fully constrained in rotation and translation at both ends. (B) The fiber subjected to illumination with $\psi_1=0^{\circ}$ and $\psi_2=90^{\circ}$ undergoes periodic flapping motion between two stable states. (C) The fiber subjected to illumination with $\psi_1=15^{\circ}$ and $\psi_2=90^{\circ}$ undergoes periodic whirling motion about its end-to-end (X) axis. (D) The lowest eigenvalue of stiffness matrix of fiber (B) during periodic flapping motion. (E) The lowest eigenvalue of stiffness matrix of fiber (C) during periodic whirling motion. (F) Illumination direction affect the frequency and pattern of motion of initially buckled fiber. In these simulations, $L_b=0.95$ and $\phi=0^{\circ}$.}
  \label{fig5}
\end{center}
\end{figure}

\paragraph{Zero twist}
We now consider the case of zero twist ($\phi=0$). The pre-stressed shape is planar and we assume that it belongs to the XY plane as shown schematically in Figure \ref{fig5}A. There are two cases. The first is the planar case where the illumination is also in the XY plane ($\psi_2=0^\circ$). This is exactly the situation studied by Gelebart {\it et al.} \cite{gelebart2017making} experimentally (though they used a strip) and through modeling by Korner {\it et al.} \cite{korner_nonlinear_2020} (though they used a quasistatic model with no inertia).  When illuminated, we observe that the fiber goes into a flapping motion as shown in Figure \ref{fig5}B coinciding with the results of Korner {\it et al.} \cite{korner_nonlinear_2020}. The second case is the non-planar case where the illumination is not in the XY plane ($\psi_2 \ne 0^\circ$).  In this case, we see in Figure \ref{fig5}C that the fiber goes into whirling motion similar to that of the case with twist.  

There is  a crucial difference between the flapping motion in the planar case and  the whirling motion in the non-planar case. Korner {\it et al.} \cite{korner_nonlinear_2020} showed that the {\it flapping} motion in the planar case is enabled by snap-through buckling {\it instability} between two distinct buckled states.  This is also shown in Figure \ref{fig5}B where the arrows trace the position of the peak during one cycle.  There are two segments where the evolution is slow as the peak/trough moves to the right (dashed horizontal arrows), and two segments where the evolution is fast
as the fiber snaps between the up and down buckled states (diagonal arrows).  The snap-through is confirmed by analyzing the stability by examining the smallest eigenvalue of the stiffness matrix in Figure \ref{fig5}D. We note that this becomes negative indicating instability.  
In contrast, the motion in {\it whirling} is always smooth and there is {\it no instability}.  This is confirmed by analyzing the stability by examining the smallest eigenvalue of the stiffness matrix in Figure \ref{fig5}E.  This is also in the case of an initially buckled and twisted fiber ($\phi \ne 0^\circ$) (see supplemental information). Figure \ref{fig5}F shows he frequency of flapping is smallter than that of whirling. 

\section{Photo-driven propulsion}
\begin{figure}[t]
\begin{center}
  \includegraphics[width=6in]{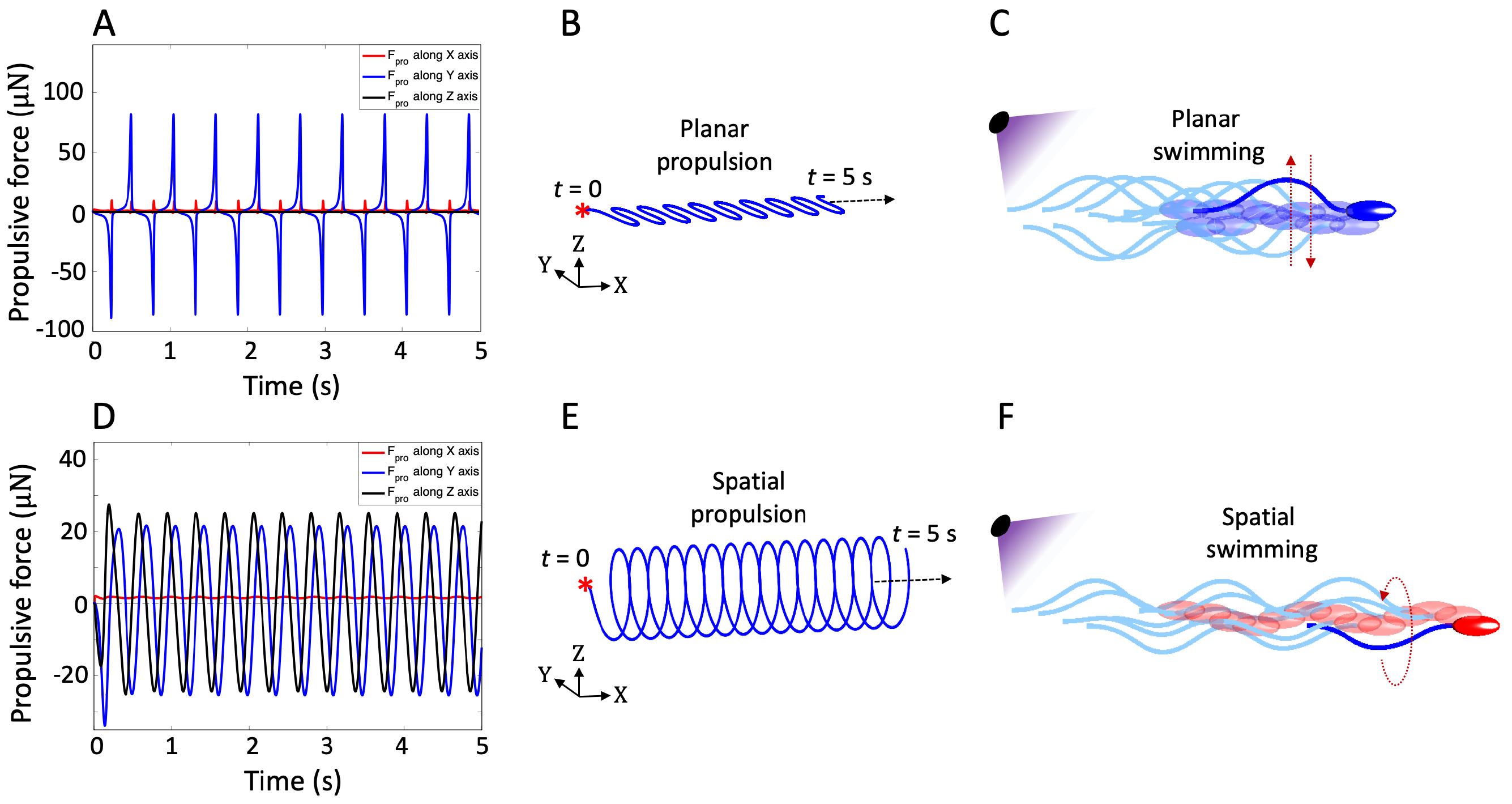}
  \caption{Photoactivation of fiber subjected to hydrodynamic drag produces a steady propulsion along X axis. For a planar fiber ($\phi=0^{\circ}$), (A) net force along Z axis remains zero, and (B) the trajectory of propulsion is sinusoidal. For a spatial fiber ($\phi \neq 0^{\circ}$), (C) the average of net forces along Y and Z axes remain zero, and (D) the trajectory of propulsion is helical. In these simulations $\Lambda=500\mathrm{W/m^2}$, $\psi_1=30^{\circ}$, and $\psi_2=0^{\circ}$.}  
\label{fig6}
\end{center}
\end{figure}

We now examine whether the photo-driven motion discussed in the previous section can be used to generate propulsion. We consider that the fiber is immersed in a fluid and therefore subject to a hydrodynamic drag force. The total drag force $\vec F_{pro}$ on the filament in laboratory frame is given by integrating the drag force per unit length over the length of the filament
\begin{equation}
\vec F_{pro} = \int_0^L \vec f_{drag} \mathrm{d}s \label{eq:11} .
\end{equation}
Further, the average propulsive force over the cycle is
\begin{equation}
\vec {\bar {F}}_{pro} = \frac{1}{T}\int_0^T  \vec F_{pro} \mathrm{d}t  \label{eq:12};
\end{equation}
where $T$ is the period.

We assume that the fiber is attached to a bead, and therefore this force propels the bead through the fluid with the velocity $\vec V = -C^{-1} \vec F_{pro}$ where $C$ is the drag coefficient for the bead and $\vec V$ is the velocity of bead in laboratory frame.  We solve the equations of the fiber to obtain the total force, and then use this force to study the motion of the bead.  We now describe results where the immersing fluid is water with  viscosity of $\mu=0.001$Pa.s, and for simplicity, we assume $C$ is identity. 

Figure \ref{fig6} shows the results for a planar fiber ($\phi=\psi_2=0^{\circ}$) with a flapping motion and a spatial fiber ($\phi \neq 0^{\circ}$) with a whirling motion.  We first consider the {\it planar} case.  Figure \ref{fig6}A shows that the propulsive force is periodic after an initial transient; this is expected since the motion and the velocity $\vec v$ of fiber are periodic as noted earlier.  There is a large propulsive force in the $Y-direction$ that is normal to the axis of the fiber; however the average of this force is zero.  The propulsive force in the Z direction normal to the plane is zero as expected.  Finally, we have spikes in the propulsive force in the X direction during the snap-through phase of the cycle; importantly, this is always in the same direction, and thus the net  propulsive force (Eq. (\ref{eq:12})) in the X direction is non-zero. All of this leads to the motion of the bead shown in Figure \ref{fig6}B. Note that it follows a sinusoidal path of cyclic excursions in the Y direction that produce no net translation, but a quasi-steady motion in the X direction.    This sinusoidal trail is reminiscent of the crawling of the nematode and C. elegance.

We then turn to the spatial case. Figure \ref{fig6}C shows that the propulsive force is periodic after an initial transient as expected.  Once again we see large forces in the Y and Z directions, but they average to zero.  However, there is a small but steady force in the X direction.  This leads to the helical or spiral motion as shown in Figure \ref{fig6}D that is reminiscent of swimming of sperm and flagellated bacteria.

\begin{figure}
\begin{center}
  \includegraphics[width=6.5in]{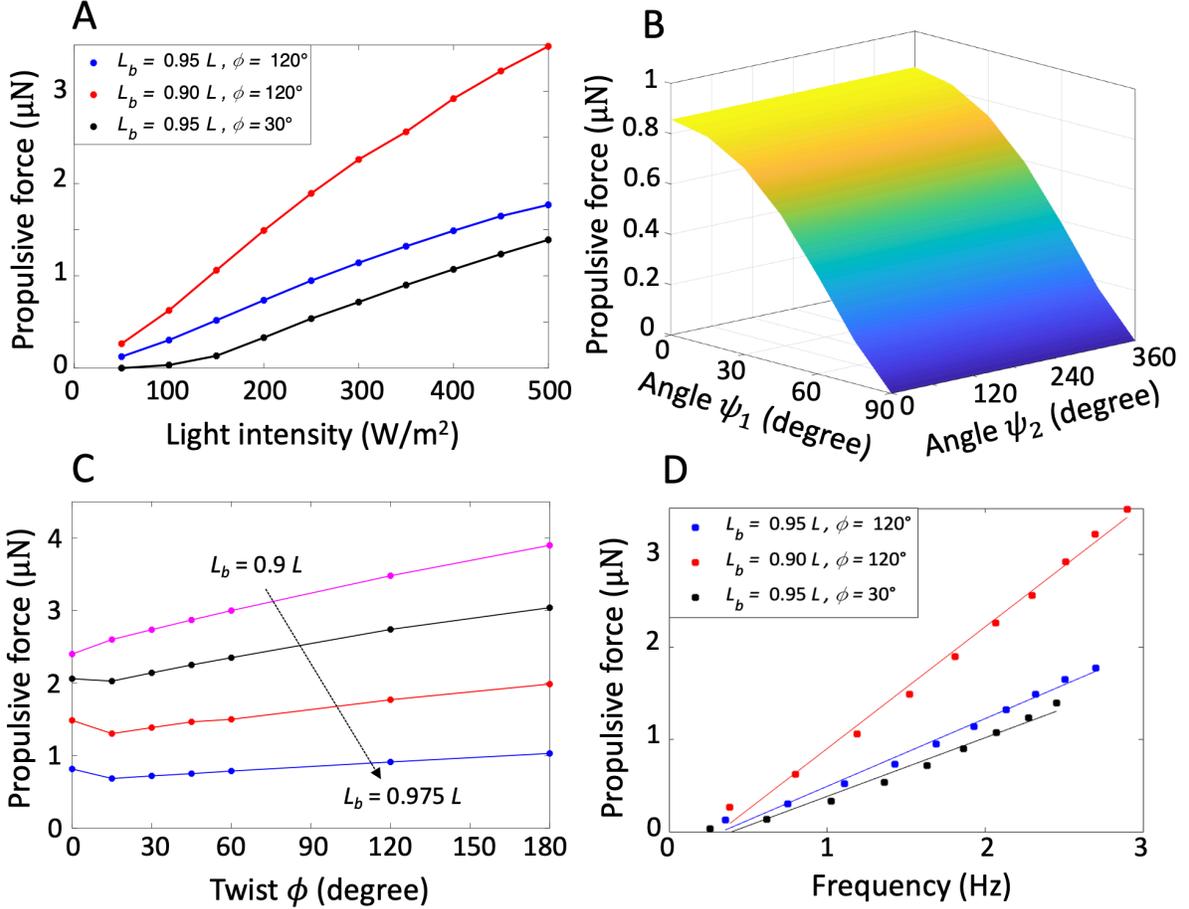}
  \caption{The effect of illumination and pre-stressed  conformation  of  fiber  on  propulsion. (A) Propulsive force as a function of light intensity, where $\psi_1=30^{\circ}$ and $\psi_2=0^{\circ}$. (B) Propulsive force as a function of illumination direction for various illumination angles $\psi_1$ and $\psi_2$, where $L_b=0.95L$ and $\phi=120^{\circ}$, $\Lambda=200\mathrm{W/m^2}$. (C) Propulsive force as a function of the initial twist $\phi=\{0^{\circ},15^{\circ},30^{\circ},45^{\circ},60^{\circ},120^{\circ},180^{\circ}\}$ and buckling $L_b=\{0.9,0.925,0.95,0.975\}L$. In these simulations $\Lambda=500\mathrm{W/m^2}$, $\psi_1=30^{\circ}$, and $\psi_2=0^{\circ}$. (D) Propulsive force vs. frequency at different illumination intensities $\Lambda=[50-500]\mathrm{W/m^2}$. }  
\label{fig7}
\end{center}
\end{figure}
Figure \ref{fig7} presents a parameter study. Figure \ref{fig7}A shows that the average propulsive force over the cycle increases with illumination. This is consistent with our finding that the frequency increases with illumination (cf. supplementary information). Figure \ref{fig7}B shows that the role of illumination direction, and this reflects the dependance of the frequency since the twist and the buckled length remain the same. Similarly, Figure \ref{fig7}C shows that propulsive force increases with increasing twist, and that this increase is greater than that can be expected from the increase of frequency alone (cf. supplementary information). Also, the propulsive force increases with decreasing $L_b$, and that this dependance is greater than that of the frequency. 

Figure \ref{fig7}D shows that the propulsive force and the frequency are proportionally correlated. This is consistent with the resistive force theory and experimental data \cite{rodenborn2013propulsion,zhong2013flow} for a rotating rigid cylindrical helix at low Reynolds number that the propulsive thrust increases linearly with respect to rotation frequency. We note that the slight deviation from a perfect linear function can correspond to the slight deformation of fiber during rotation while its overall shape remains almost constant (cf. Figure \ref{fig2}).


\paragraph{Acknoweledgment}
We gratefully acknowledge the support of the US Office of Naval Research through Multi-investigator University Research Initiative Grant ONR N00014-18-1-2624.


\newpage
\setcounter{equation}{0}
\setcounter{section}{0}
\setcounter{figure}{0}
\setcounter{page}{1}

\begin{center}
{\Large \bf  Supplementary Material}\\ 
\vspace{0.05in}

Light-induced swirling and locomotion\\
Ameneh Maghsoodi and Kaushik Bhattacharya
\end{center}
\vspace{.3in}

\section{Photomechanical model of Azo-LCE fibers}
\begin{figure}
\begin{center}
  \includegraphics[width=5in]{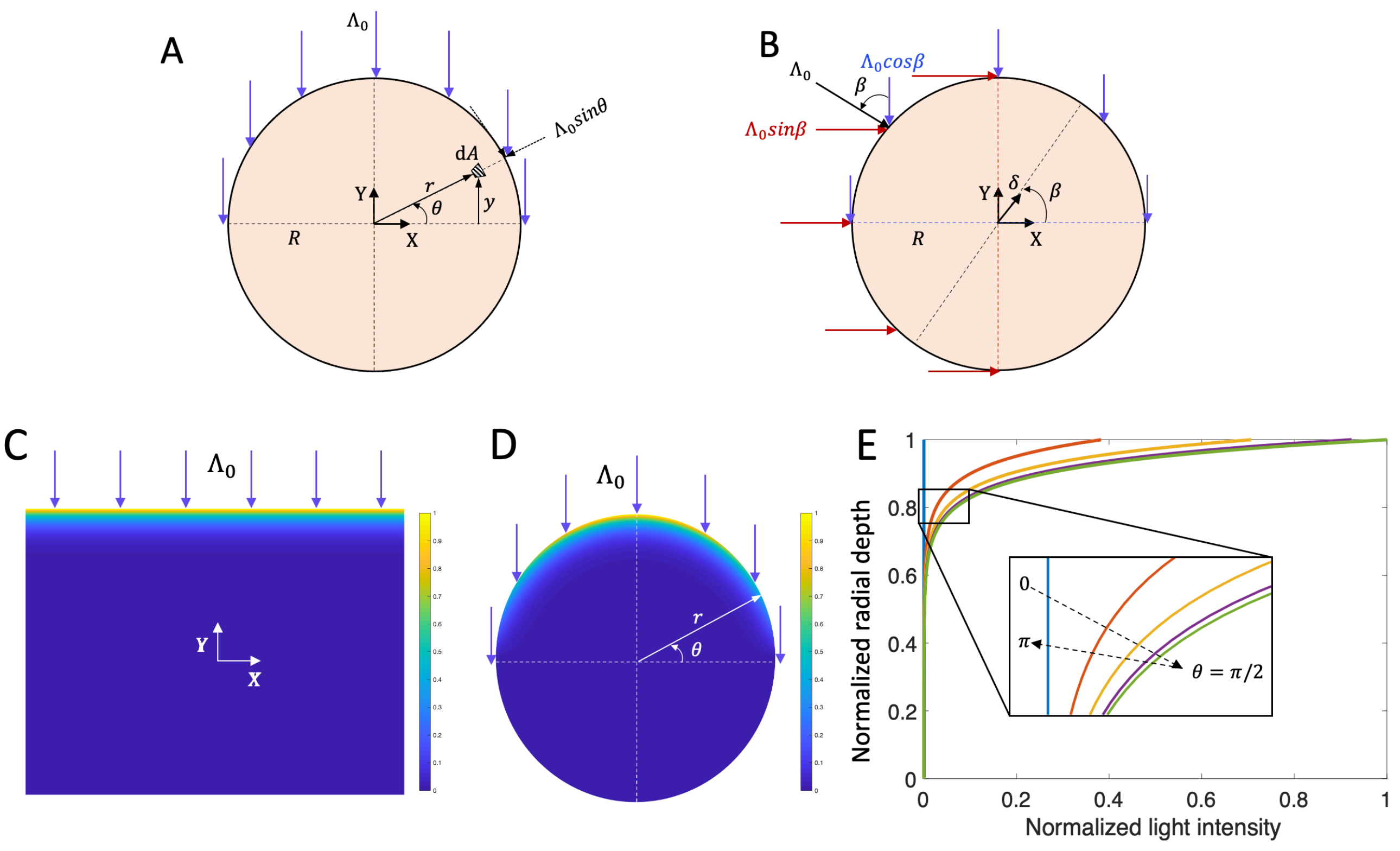}
  \caption{Illumination of a circular cross-sectional fiber where illumination (A) is along principal axis, and (B) makes arbitrary angle $\beta$ with principal axis Y. (C) Profile of light intensity at a rectangular cross-section. (D) Profile of light intensity at a circular cross-section. (E) The light intensity penetrating a circular cross-section varies with both radial depth and angle.}
  \label{figS1}
\end{center}
\end{figure}

Consider a long, thin azo-LCE fiber possessing circular cross-section. Under illumination, Azobenzene molecules absorb photons and transform between linear $trans$ to bent $cis$ configurations due to light absorption and thermal decay. Suppose for the moment that the fiber is subjected to an illumination with intensity $\Lambda_0$. As a simple case, assume that illumination is along axis Y as shown in Figure \ref{figS1}A. As the light diffracts into the cross-section, it is absorbed by the photochrome molecules and attenuates. In this work, we assume shallow penetration governed by Beer's law. Recall that, according to Beer's law, when a light $\Lambda_0$ impinges on the flat surface of a strip with thickness $H$, the intensity decreases with depth such that the intensity at any point $y$ on the cross-section is given by

\begin{equation}
\Lambda(s,t,y)=\Lambda_0(s,t) \mathrm{exp}\left(-\frac{H/2-y}{d}\right)=\Lambda_0(s,t) f(y) , y \in [{0},H/2] \label{eq:s1}.
\end{equation}
where $d$ is the penetration depth. Similarly, for a fiber with circular cross-section shown in Figure \ref{figS1}A

\begin{equation}
\Lambda(s,t,r,\theta)=\Lambda_0(s,t) sin\theta  \mathrm{exp}\left(-\frac{R-r}{d}\right)=\Lambda_0(s,t) f(r,\theta) , r \in [{0},R], \theta \in [{0},\pi] \label{eq:s2}.
\end{equation}
where function $f\in [{0},1]$. 

Figure \ref{figS1}C-E compares the light intensity profile for a circular and a rectangular cross-sections using (\ref{eq:s1}) and (\ref{eq:s2}). As shown in Figure \ref{figS1}C, the light intensity passing through a rectangular cross section decreases (exponentially) with depth $H$. However, Figure \ref{figS1}D-E for a circular cross-section show that the light intensity depends on the depth $r$ as well as the angle between incident light and radial direction; where the maximum light intensity corresponds to $\theta=\pi/2$ and $r=R$. Note that the light intensity at any angle $\theta$ decreases exponentially with radial depth (cf. (\ref{eq:s2})). This demonstrates that the fiber cross sectional shape affects the profile of light intensity.

The photochromes in the {\it trans} state absorb light and transform into the {\it cis} state which can thermally relax back to the {\it trans} state. Therefore, the number density $n_c$ of {\it cis} molecules along cross-section evolves according to 
\begin{equation}
\tau \dot n_c = -n_c +\alpha \Lambda  = -n_c +\alpha \Lambda_0 f \label{eq:s3},
\end{equation}
where the dot represents the material time derivative, $\tau$ is thermal relaxation time (or $cis$ lifetime), and $\alpha$ is a constant depending on material properties and forward-backward reaction rates.  This in turn introduces a spontaneous axial strain $\varepsilon_c = - \lambda n_c$ where $\lambda$ is a constant for proportionality with $\lambda>0$ corresponds to a contraction. Consequently, the light-induced bending at the circular cross-section of the fiber 
\begin{equation}
q(s,t)=B(\kappa(s,t)-\kappa_0(s,t)) \label{eq:s4},
\end{equation}
where
\begin{equation}
\kappa_0(s,t)=\kappa_{in}(s) + \int_{0}^{2\pi} \int_{0}^{R} \left(\frac{-\lambda}{I_c}\right)n_c(s,t,r,\theta) r^2 sin\theta \mathrm{d}r \mathrm{d}\theta \label{eq:s5},
\end{equation}
In (\ref{eq:s4}), $\kappa$ and $\kappa_0$ denote the curvature and spontaneous stress-free curvature of the fiber, respectively. The constant $B=EI_c$ is the bending stiffness with the second moments of area $I_c=\pi R^4/4$ for a circular cross-section. Combining (\ref{eq:s3})-(\ref{eq:s5}), 
\begin{equation}
\tau \dot\kappa_0(s,t)+(\kappa_0(s,t)-\kappa_{in}(s))= \gamma \Lambda_0(s,t), \label{eq:s6} 
\end{equation}
where
\begin{equation}
\gamma=\left(\frac{-\alpha \lambda}{I_c}\right) \int_{0}^{\pi} \int_{0}^{R} (r sin\theta)^2 \mathrm{exp}\left(-\frac{R-r}{d}\right)\mathrm{d}r \mathrm{d}\theta  \label{eq:s7}.
\end{equation}

Now, in a general case, assume that a steady illumination $\Lambda_0$ makes angle $\beta$ with principal axis Y as illustrated in Figure \ref{figS1}B. Illumination $\Lambda_0$ makes bending about axis $\delta$ such that
\begin{equation}
\kappa_{0X}(s,t)= \kappa_{0\delta}(s,t)cos(\beta),\label{eq:s8}
\end{equation}
\begin{equation}
\kappa_{0Y}(s,t)= \kappa_{0\delta}(s,t)sin(\beta),\label{eq:s9}
\end{equation}
On the other hand, illumination $\vec\Lambda_0=\begin{pmatrix} \Lambda_0 sin(\beta) \\\Lambda_0 cos(\beta)\end{pmatrix}$.  Therefore, using (\ref{eq:s6})
\begin{equation}
\tau \frac{\partial \kappa_{0X}(s,t)}{\partial t}+(\kappa_{0X}(s,t)-\kappa_{inX}(s))= \gamma \Lambda_0(s,t) cos(\beta), \label{eq:s10}
\end{equation}
\begin{equation}
\tau \frac{\partial \kappa_{0Y}(s,t)}{\partial t}+(\kappa_{0Y}(s,t)-\kappa_{inY}(s))= \gamma \Lambda_0(s,t) sin(\beta), \label{eq:s11}
\end{equation}
Combining (\ref{eq:s6}-\ref{eq:s11}) concludes that photoactuaction by the steady illumination $\Lambda_0$ can be written as the sum of two photoactuactions $\Lambda_0 sin(\beta)$ and $\Lambda_0 cos(\beta)$ along principal axes X and Y, respectively. 

\section{Stability analysis}
To examine stability of fiber for the dynamic equilibrium configurations, we calculate the second variation of total potential energy, $\delta^2 \Pi$. If $\delta^2 \Pi > 0$, the potential energy is minimized at the equilibrium configuration, indicating that the configuration is stable. To derive the stability criteria, consider the potential energy function in the form
\begin{eqnarray} 
\Pi=\int_{0}^{1} f\left(s,Y(s),Y'(s)\right) ds  
\label{eq:s12}
\end{eqnarray} 
where $Y=[y_1 , y_2 ,..., y_N]^T$ and $Y'=[y'_1 , y'_2 ,..., y'_N]^T$.
To calculate the second variation of energy, we consider small variations around equilibrium configuration $Y^*$ in the form
\begin{eqnarray} 
Y(s)=Y^*(s) + \epsilon \eta(s) 
\label{eq:s13}
\end{eqnarray} 
where $\epsilon$ is an arbitrary constant independent of $s$ and $Y$, and $\eta(s)=[\eta_1 , \eta_2 ,..., \eta_N]^T$ contains arbitrary functions which are independent of $\epsilon$ and satisfy boundary conditions. We differentiate $\Pi$ twice with respect to $\epsilon$ and set $\epsilon$ to zero
\begin{eqnarray} 
\delta^2 \Pi=\int_{0}^{1}  \left (\eta^T \frac{\partial^2 f}{\partial Y \partial Y}\eta + 2 \eta^T \frac{\partial^2 f}{\partial Y \partial Y'} \eta' +  \eta'^T \frac{\partial^2 f}{\partial Y' \partial Y'} \eta' \right) ds  
\label{eq:s14}
\end{eqnarray}
From Rayleigh-Ritz approach, we assume $\eta_i=\Sigma^{n}_{j=1} a_{ij} \phi_j$, $i=1,2,.., N$. In doing so, finding the minimum of second variation of energy subject to the constraint that $\int_{0}^{1} \eta^T \eta ds=1$ becomes (by Lagrange multiplier $\lambda$) an eigenvalue problem of finding $\lambda$. The positive eigenvalues imply that the configuration is stable. For a twisted-buckled rod that is clamped at one end and subjects to force $\vec F$ and moment $\vec M$ at the other end 
\begin{eqnarray} 
\Pi=\int_{0}^{1}  \left(\frac{\mathbf B}{2} \left(\frac{\partial \vec \theta}{\partial s} - \vec \kappa_0 \right) ^2 - \vec F. \vec r' - \vec M.\frac{\partial \vec \theta}{\partial s} \right) ds 
\label{eq:s15}
\end{eqnarray} 
Where $Y=\vec \theta(s)= [\theta_1(s) , \theta_2(s) ,\theta_3(s)]^T$, and $\vec r'=\mathbf {R}(\vec \theta).\vec a_3$ in which $\mathbf {R}$ is the rotation matrix from local frame to inertial frame.

\begin{figure}
\begin{center}
  \includegraphics[scale=0.45]{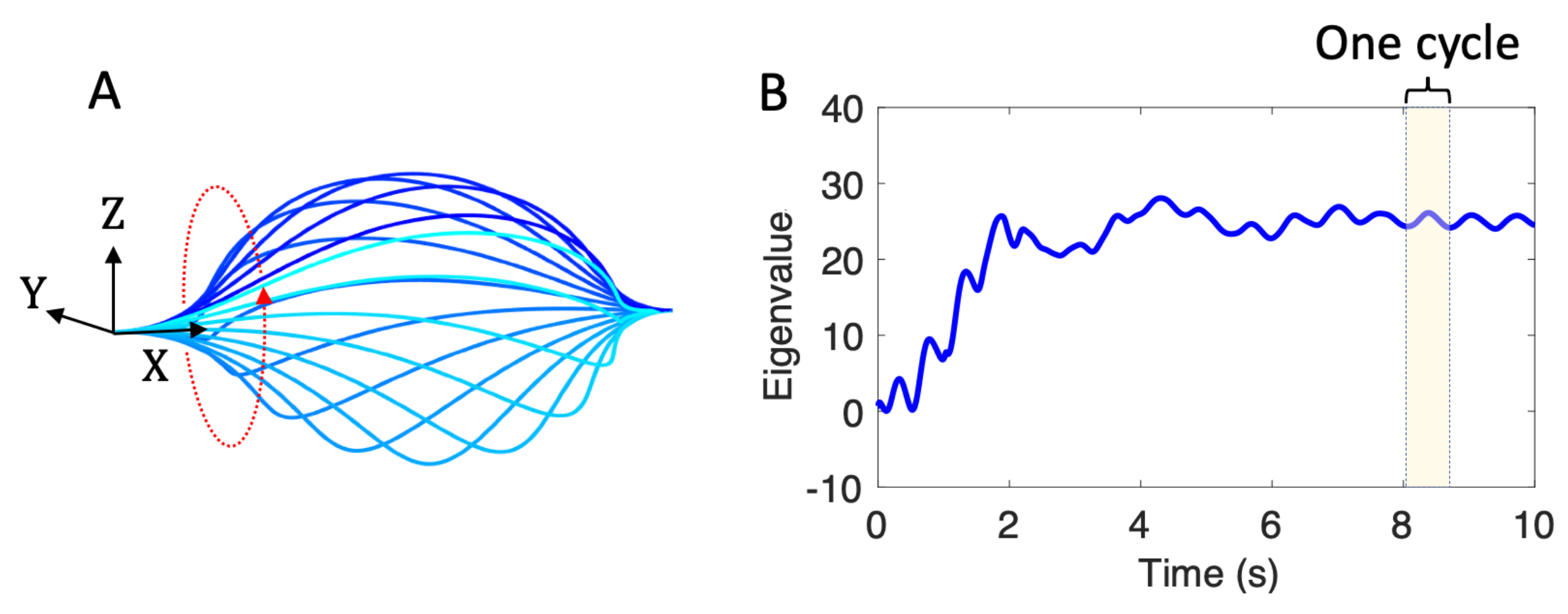}
  \caption{(A) Whirling motion of pre-stressed buckled-twisted fiber about its helical axis. (B) The lowest eigenvalue of stiffness matrix during periodic whirling motion remains positive.}
  \label{figS2}
\end{center}
\end{figure}

As shown in Figure \ref{figS2}A, a pre-stressed buckled-twisted fiber undergoes a periodic whirling motion under steady illumination. The motion in {\it whirling} is always smooth and there is {\it no instability}. This is confirmed by analyzing the stability by examining the smallest eigenvalue of the stiffness matrix in Figure \ref{figS2}B. The lowest eigenvalue is always positive.

\section{Parameter Study}

\paragraph{The effect of illumination intensity and direction}

\begin{figure}
\begin{center}
  \includegraphics[width=6.5in]{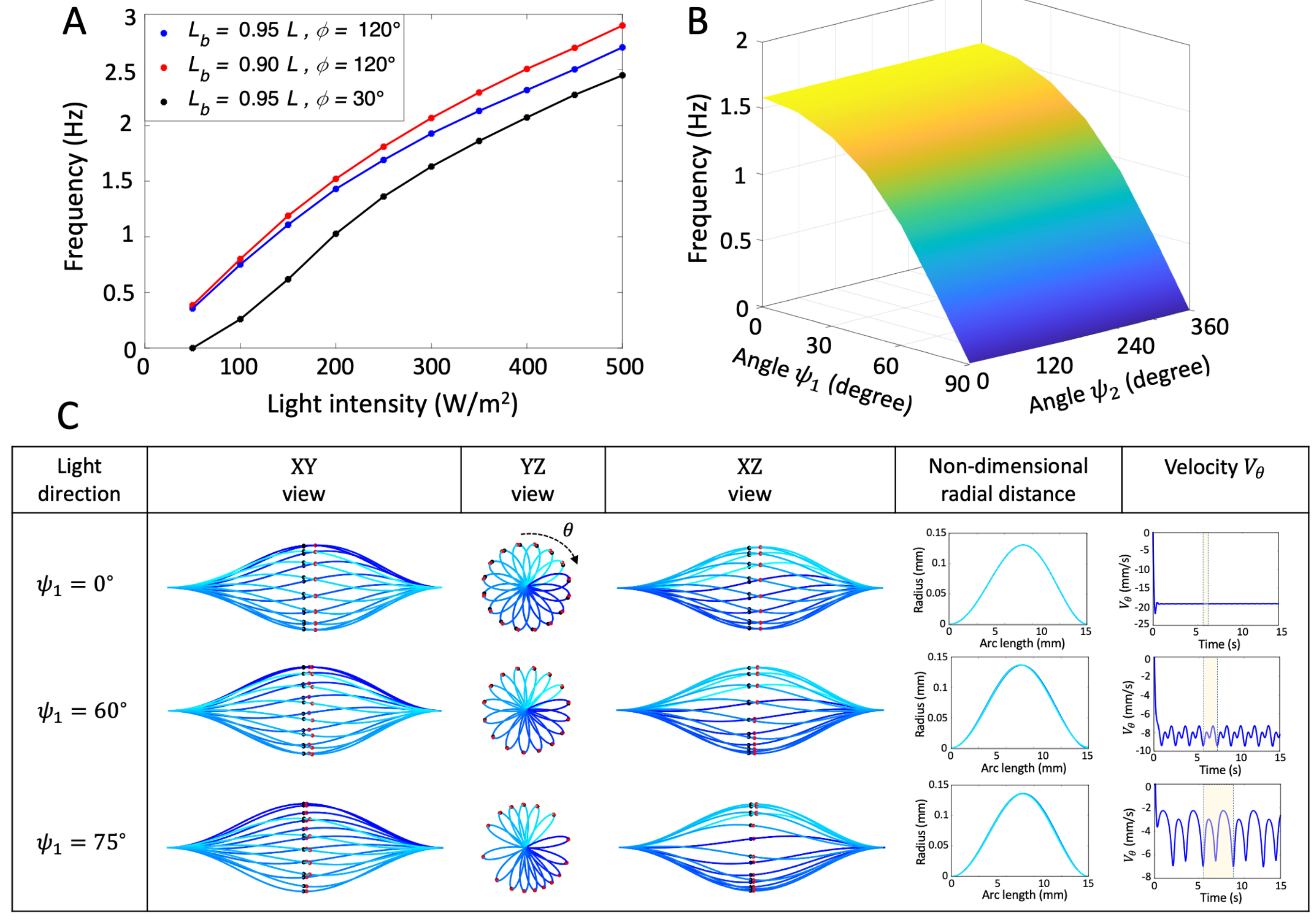}
  \caption{Influence of illumination on the frequency and pattern of oscillations. (A) Frequency of oscillations as a function of light intensity, where $\psi_1=30^{\circ}$ and $\psi_2=0^{\circ}$. (B) Frequency of oscillations as a function of illumination direction for various illumination angles $\psi_1$ and $\psi_2$, where $L_b=0.95L$ and $\phi=120^{\circ}$, $\Lambda=200\mathrm{W/m^2}$. (C) Comparison of conformational change of fiber during periodic oscillations at three different illumination directions. Red dots refer to maximum radius of fiber and black dots refer to mid-point.}
  \label{fig3}
\end{center}
\end{figure}

Figure \ref{fig3} compares the frequency $f$ and pattern of oscillations over a range of values of illumination intensity $\Lambda$ and angles $\psi_1$ and $\psi_2$.   One needs a critical intensity before one has periodic motion (Figure \ref{fig3}A); below this critical intensity, the fiber deforms and attains a photo-stationary state.   Beyond this critical intensity, the frequency increases with the light intensity: this is consistent with the fact that greater intensity leads to a higher rate of increase in the photo-induced strain and spontaneous curvature. 

The angle $\psi_1$ between the direction of light propagation and the X-axis (the line joining the end-points) has a significant effect on the motion.  When the illumination is perpendicular to the X-axis, then the fiber reaches a photo-stationary state (Figure \ref{fig3}B).  The frequency increases as $\psi_1$ decreases and reaches a peak when the illumination is parallel to X-axis. There is symmetry about $\psi_1 = 90^\circ$; the illumination at angle $-\psi_1$ drives fiber with the same velocity but opposite direction of illumination at angle $\psi_1$.

Figure \ref{fig3}C shows that this angle also affects the shape of the rod, and the rotary motion becomes less steady as the illumination is perpendicular and the more steady as it is parallel to the X-axis.   The angle $\psi_2$ which describes orientation in the YZ plane has no impact, consistent with almost rotary motion.

\paragraph{The effect of pre-stressed conformation of fiber}

\begin{figure}
\begin{center}
  \includegraphics[width=6.5in]{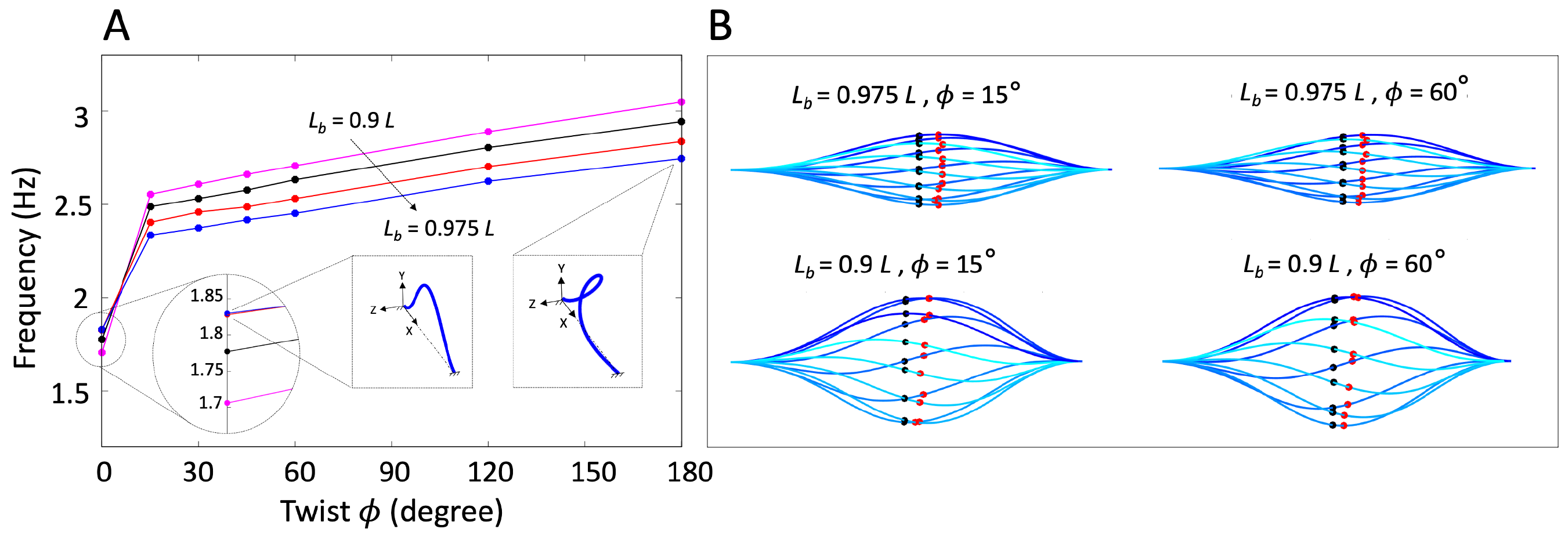}
  \caption{The effect of pre-stressed conformation of fiber on the oscillations. (A) Frequency of periodic oscillations as a function of the initial twist $\phi=\{0^{\circ},15^{\circ},30^{\circ},45^{\circ},60^{\circ},120^{\circ},180^{\circ}\}$ and buckling $L_b=\{0.9,0.925,0.95,0.975\}L$. The inset compares the conformation of planar ($\phi=0^{\circ}$) fiber and spatial ($\phi \neq 0^{\circ}$) fiber. (B) Comparison of conformational change of fiber during periodic oscillations under four different pre-stressed conformations. Red dots refer to maximum radius of fiber and black dots refer to mid-point. In these simulations, $\Lambda=500\mathrm{W/m^2}$ with $\psi_1=30^{\circ}$ and $\psi_2=0^{\circ}$.} 
\label{fig4}
\end{center}
\end{figure}

Figure \ref{fig4} examines the frequency as a function of the pre-stressed conformation, both the buckled length $L_b$ and the angle of twist $\phi$ (with  $\Lambda=500\mathrm{W/m^2}$ with $\psi_1=30^{\circ}$ and $\psi_2=0^{\circ}$).    The figure shows that the case of no twist, $\phi = 0$, is distinct and so we defer its discussion. Once twisted, Figure \ref{fig4}A shows that the frequency increases almost linearly but only slightly with increasing twist. Increasing the buckling (i.e., reducing $L_b$) also increases the frequency.  Figure \ref{fig4}B shows that the shape changes with a larger bulge with reduced $L_b$.

\begin{figure}
\begin{center}
  \includegraphics[width=6.5in]{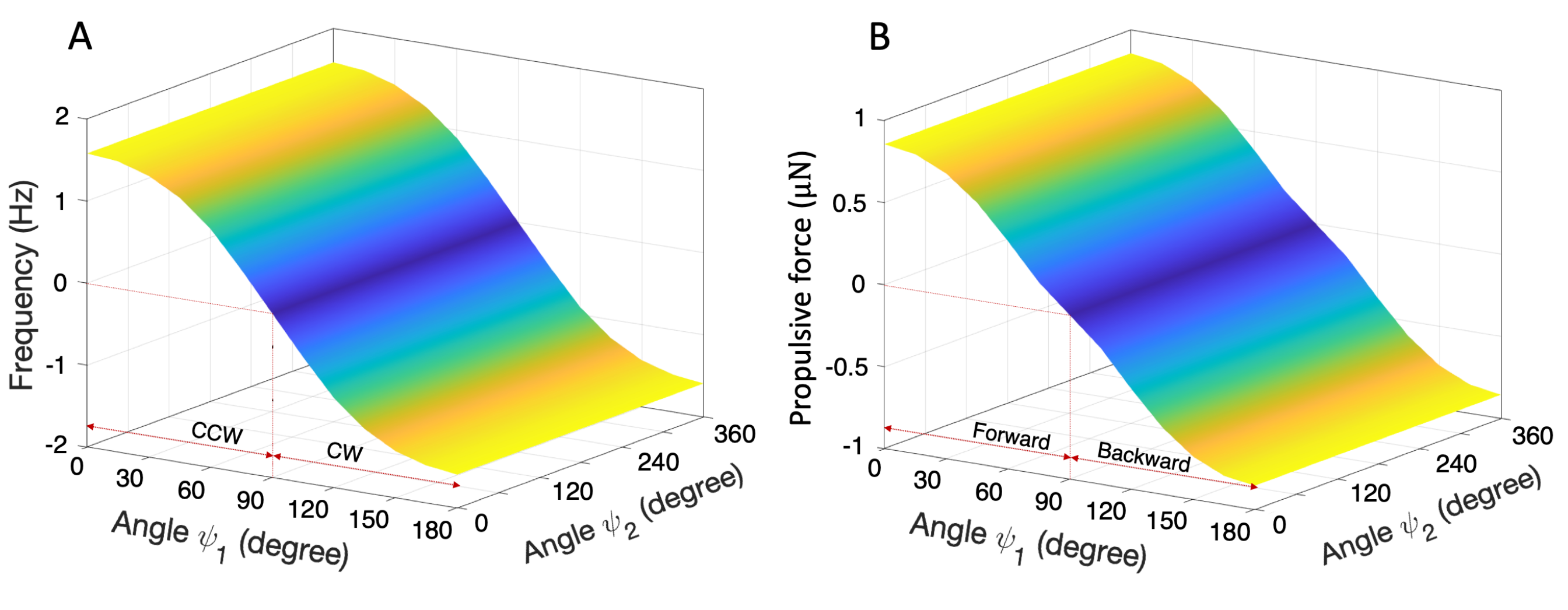}
  \caption{(A) Frequency of oscillations as a function of illumination direction. (B) Propulsion as a function of illumination direction. In this simulations, $L_b=0.95L$ and $\phi=120^{\circ}$, and $\Lambda=200\mathrm{W/m^2}$. The frequency and propulsive force are symmetric with respect to YZ plane ($\psi_1=90^{\circ}$)} 
  \label{figS3}
\end{center}
\end{figure}

\paragraph{Influence of illumination direction on the frequency and propulsion}
As shown in Figure \ref{figS3}, the whirling motion of a pre-stressed twisted-buckled fiber is symmetric with respect to the YZ plane perpendicular to the helical axis if fiber. Note that the amplitude of frequency and propulsive force are the same but the directions are the opposite with respect to YZ plane.  

\paragraph{The effect of cross-sectional geometry}
We now briefly examine whether the cross-section of fiber affects the photo-driven motion. In photomechanical model of fiber, the function $f_1(\vec y)$ depends on both geometrical and material properties of fiber. Examples of function $f_1(\vec y)$ are presented and compared for circular and rectangular cross section in the supplementary material. Figure \ref{fig8} compares the photomechanical response of the fibers with square and rectangular cross sections where the illumination is along X-axis from left to right. From Figure \ref{fig8}A-C for a square cross-sectional fiber, the motion is periodic. The fiber rotates uniformly about its helical axis with an almost constant shape; this is similar to that of circular cross-sectional fiber explained in previous sections. 

Figure \ref{fig8}D-F shows the photomechanical response of a rectangular cross-sectional fiber. The photo-driven motion is periodic; however, the pattern of deformation is significantly different from circular cross-sectional fiber. The fiber rotates about its axis, however, the rotation velocity is largely nonuniform during the cycle as observed in Figure \ref{fig8}D. From Figure \ref{fig8}E-F, the radial distance of fiber relative to rotation axis significantly changes. Also the maximum radius (Red dots) translates along X-direction while rotating about X-axis. These results concludes that the fiber undergoes a {\it coupled flapping-whirling} motion about its axis. Future analysis and parametric study on rectangular cross-section can reveal the contribution of bending and torsional elasticity and aspect ratio on the coupled flapping-whirling motion.         
\begin{figure}[t]
\begin{center}
  \includegraphics[width=6.5in]{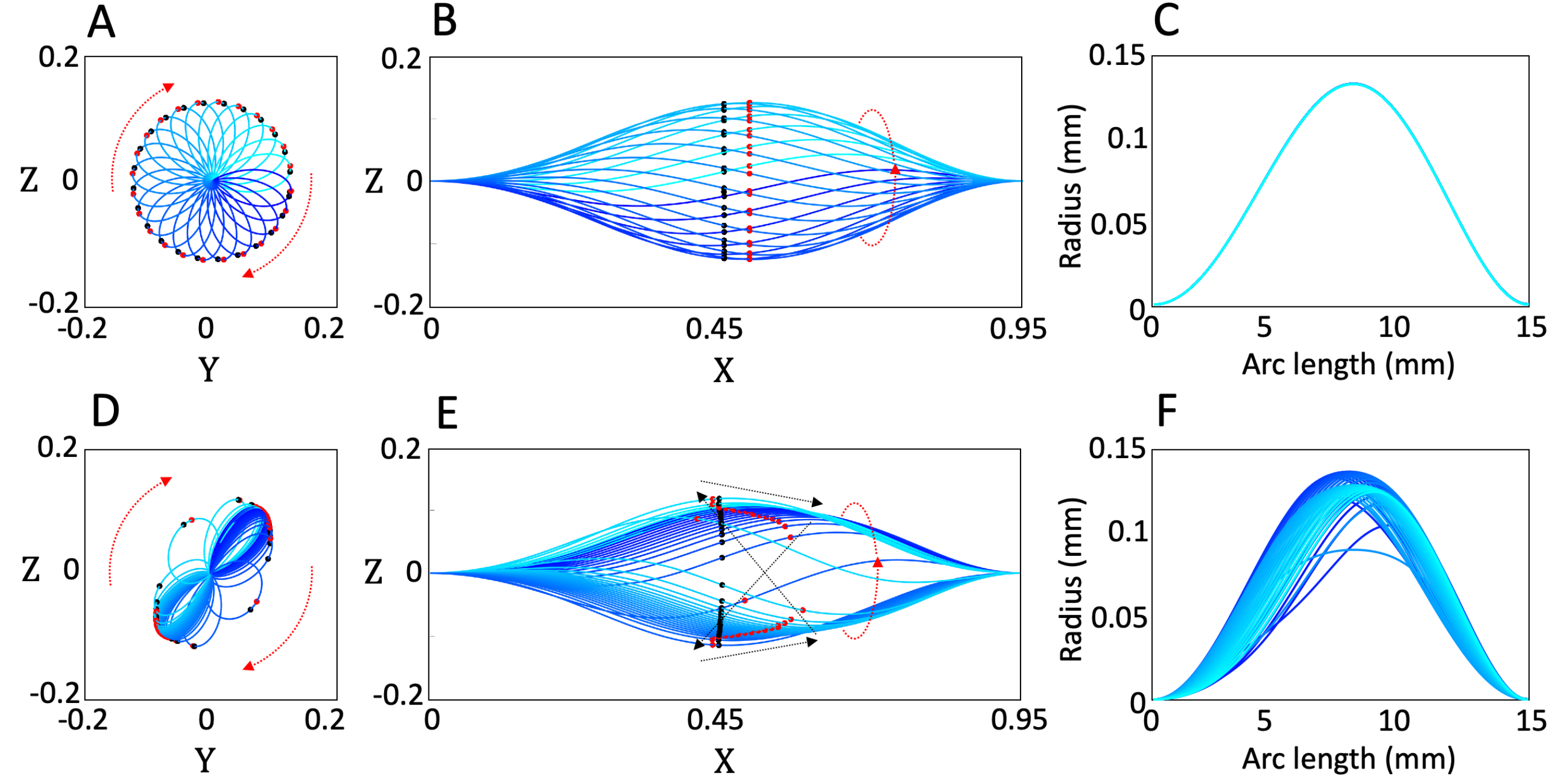}
  \caption{The effect of fiber cross-section on photo-driven motion. Comparison of conformational change of fiber with (A-C) square cross-section with width and thickness of 15 $mu$ m, and (D-F) rectangular cross-section with width of 15 $mu$ m and thickness of 22 $mu$ m. Note that (C) and (F) present the radial distance of fiber from its end-to-end axis. Also, red arrows indicate the direction of rotation, black arrows in (E) shows the flapping trajectory, the red dots refer to maximum radial distance, and black dots indicate the middle of fiber at each time. In these simulations $L_b=0.95L$, $\phi=\{120^{\circ}$, $\Lambda=500\mathrm{W/m^2}$, $\psi_1=0^{\circ}$, and $\psi_2=0^{\circ}$.}  
\label{fig8}
\end{center}
\end{figure}

\section{Animations}

\paragraph{Planarpropulsion.mp4}  Shows the motion of a bead attached to a buckled (but not twisted) fiber illuminated along the buckled plane.

\paragraph{Spatialpropulsion.mp4} Shows the motion of a bead attached to a buckled (but not twisted) fiber illuminated at an angle to the buckled plane.

\end{document}